\documentstyle[12pt]{article}

\def\be{\begin{equation}}
\def\ee{\end{equation}}
\def\bea{\begin{eqnarray}}
\def\eea{\end{eqnarray}}

\def\p{\psi_{-}}

\begin{document}
\vspace*{2cm}
\begin{center}
{\large\bf DUALITY AND CANONICAL TRANSFORMATIONS}\footnote{Talk
given at the Argonne Duality Institute, June 27-July 12, 1996.
To appear in the
e-proceedings.}
\vskip 1.5 cm
{\bf Y. Lozano}\footnote{\tt Y.Lozano@fys.ruu.nl}
\vskip 0.05cm
Inst. for Theoretical Physics,
Utrecht University,
3508 TA Utrecht, The Netherlands
\end{center}

\date{ }
\setcounter{page}{0} \pagestyle{empty}
\thispagestyle{empty}
\vskip 3 cm
\begin{abstract}
We present a brief review on 
the canonical transformation description of some
duality symmetries
in string and gauge theories. In particular,
we consider abelian and non-abelian T-dualities 
in closed and open string theories as well as
S-duality in abelian and non-abelian 
non-supersymmetric gauge
theories.
\end{abstract}
\vfill
\begin{flushleft}
THU-96/36\\
hep-th/9610024\\
October 1996
\end{flushleft}
\newpage\pagestyle{plain}

\def\theequation{\thesection . \arabic{equation}}

\tableofcontents

\section{T-Duality in String Theory}
\setcounter{equation}{0}

Some suggestions had been made in the literature pointing
(at least in the simplified situation when all backgrounds
are constant or depending only on time)
towards an understanding of T-duality as particular instances
of canonical transformations \cite{GRVMV}.
This idea works well when the backgrounds have an abelian
isometry \cite{AAGL2}, laying duality on a simpler setting
than before (see for instance \cite{GPR} for a review
on previous formulations).
Esentially the canonical transformation provides the non-local
change of variables identified as world-sheet T-duality,
which reduces to
$d\rightarrow *d$ for toroidal compactifications. 
Concerning non-abelian duality it is also possible to give
a formulation in terms of canonical transformations when
the isometry acts without isotropy \cite{CZ1,L1,AL,KS1,S}.
In this case the canonical formulation allows to define the
dual theory in arbitrary genus Riemann surfaces, what was not known
within the original gauging procedure formulation \cite{DLOQ}.

There remain however some open problems within this formulation
of duality. For the dual theory to be conformally invariant
the dilaton must transform as 
${\tilde \phi}=\phi-\frac12 \log{R^2}$ 
(for toroidal compactifications,
or a straightforward generalization for abelian or non-abelian
isometries) \cite{BU,AO}. A general argument justifying this
transformation in phase space is not available, although we believe
it should be along the lines of how the modular anomaly appears
in abelian gauge theories, as we will discuss later in
the text.
Also, the explicit canonical transformation formulation of 
non-abelian
duality for isotropic isometries is not known. There are 
results in the literature \cite{KS2,S,KS3} proving that the initial
and dual theories are indeed canonically equivalent but 
the non-local mapping generalizing $d\rightarrow *d$ to these
transformations has still not been found.

\subsection{Abelian duality}

It is by now well known that abelian T-duality in sigma
models is the result of a simple canonical
transformation in the phase space of the theory \cite{AAGL2}. 
Written
in configuration space variables this transformation
generalizes the duality mapping
\be
\label{1.1}
\partial_+ x=-\frac{1}{R^2}\partial_+{\tilde x},
\qquad
\partial_- x= \frac{1}{R^2}\partial_-{\tilde x}
\ee
of toroidal compactifications to general backgrounds with abelian
isometries.
Given a general $d$ dimensional $\sigma$-model with 
an abelian isometry represented by
translations of a $\theta$-coordinate:
\be
\label{ab1}
L=\frac12 g_{00} (\dot{\theta}^2-\theta^\prime\,^2)+(\dot{\theta}+
\theta^\prime)j_-+(\dot{\theta}-\theta^\prime)j_++\frac12 V,
\ee
where:
\bea
\label{ab2}
&&j_\pm=\frac12 k_\alpha^\mp\partial_\pm x^\alpha, \qquad
\alpha=1,\dots,d-1 \nonumber\\
&&k_\alpha^\pm=g_{0\alpha}\pm b_{0\alpha} \nonumber\\
&&V=(g_{\alpha\beta}+b_{\alpha\beta})
\partial_+x^\alpha\partial_-x^\beta,
\eea
the dual with respect to this isometry can be obtained by
performing the canonical transformation: 
\be
\label{ab3}
p_\theta=-{\tilde \theta}^\prime \qquad
p_{{\tilde \theta}}=-\theta^\prime
\ee
in the phase space of the theory or equivalently, 
the non-local mapping:
\bea
\label{ab33}
&&\partial_+\theta=-({\tilde g}_{00}\partial_+{\tilde \theta}+
{\tilde k}^-_\alpha\partial_+{\tilde x}^\alpha)\nonumber\\
&&\partial_-\theta={\tilde g}_{00}\partial_-{\tilde \theta}+
{\tilde k}^+_\alpha\partial_-{\tilde x}^\alpha
\eea
in configuration space.
The dual backgrounds are then given by Buscher's formulas
\cite{BU}.
(\ref{ab3}) is generated from the type I generating functional:
\be
\label{ab4}
{\cal F}=\int_{D,\partial D=S^1} d{\tilde \theta}\wedge d\theta=
\frac12 \oint_{S^1}(\theta^\prime {\tilde \theta}-\theta 
{\tilde \theta}^\prime)d\sigma,
\ee
which can be easily derived from the gauging 
procedure to abelian T-duality \cite{RV}.
It is easy to see that in this
approach the initial and dual Lagrangians are equivalent up to a
total derivative:
\be
\label{pro1}
\int_\Sigma d{\tilde \theta}\wedge d\theta,
\ee
usually neglected in closed strings. 
However under a canonical
transformation $\{q^i,p_i\}\rightarrow \{Q^i,P_i\}$
the initial and canonically transformed Hamiltonians
verify:
\be
\label{pro2}
\dot{q}^ip_i-H(q^i,p_i)=\dot{Q}^iP_i-
{\tilde H}(Q^i,P_i)+
\frac{d{\cal F}}{dt}
\ee
where ${\cal F}$ is the generating functional, such that
$H={\tilde H}$ if and only if (assuming ${\cal F}$ is type I 
and does 
not depend explicitly on time):
\be
\label{pro3}
\frac{\partial{\cal F}}{\partial q^i}=p_i, \qquad
\frac{\partial{\cal F}}{\partial Q^i}=-P_i.
\ee
It is then
straightforward to see that the total derivative
(\ref{pro1}) yields the generating functional and
corresponding
canonical transformation (\ref{ab4}) and (\ref{ab3}).

${\cal F}$ being linear in $\theta$ and ${\tilde \theta}$ implies
that the classical canonical transformation (\ref{ab3}) is
also valid quantum mechanically (as explained in \cite{GC})
and we can write the relation:
\be
\label{ab5}
\psi_k[{\tilde \theta}(\sigma)]=N(k)\int {\cal D}\theta (\sigma)
e^{iF[{\tilde \theta},\theta (\sigma)]}\phi_k [\theta(\sigma)]
\ee
between the corresponding Hilbert spaces. Here
$\psi_k[{\tilde \theta}]$ and $\phi_k[\theta]$ are chosen as
eigenstates
corresponding to the same eigenvalue of the respective 
Hamiltonians and
$N(k)$ is a normalization factor introduced to insure the proper
normalization of the dual wave functions.
{}From (\ref{ab5}) we can learn about the
multivaluedness and periods of the dual variables. 
Since $\theta$ is periodic,
$\phi_k(\theta +a)=\phi_k (\theta)$
implies for ${\tilde \theta}$: ${\tilde \theta}(\sigma +2\pi)-
{\tilde \theta}(\sigma)=4\pi /a$, which means that 
${\tilde \theta}$ must
live in the dual lattice of $\theta$. 

In this formulation
duality gets very simple conceptually. We can also learn
how it can be applied to arbitrary genus Riemann
surfaces, because the state $\phi_k[\theta (\sigma)]$ could be the
state obtained from integrating the original theory on an arbitrary
Riemann surface with boundary. It is also clear that the arguments
generalize straightforwardly when we have several commuting 
isometries. 
The behavior of currents not commuting with those used to
implement duality can also be clarified. In the case of
WZW models it becomes rather simple to prove that the full
duality group is given by $Aut(G)_L\times Aut(G)_R$, 
both inner and outer,
where
$L,R$ refer to the left and right currents
on the model
with group $G$. Due to the chiral
conservation of the currents in this case, the canonical
transformation leads to a local expression for the dual
currents. In the case where the currents are not chirally
conserved, then those currents associated to symmetries
not commuting with the one used to perform duality
become generically non-local in the dual theory and this
is why they are not manifest in the dual Lagrangian.

We can say that the canonical transformation 
is a ``minimal'' approach in the sense that no extraneous
structure (1-forms in Buscher's approach or fake gauge
fields in Rocek and Verlinde's) has to be introduced,
and all standard results in the abelian case (and more)
are easily recovered using it.
We should mention however that renormalization effects still 
need to be considered in
order to prove the quantum equivalence between the two theories
and there are in fact some results in the literature
showing that they give
corrections to Buscher's backgrounds \cite{correc}.
We need to reproduce as well the dilaton shift within the
canonical transformation approach. 
Consider a constant
toroidal background of radius $R$. The measure in configuration
space is given by ${\cal D}\theta {\rm det}R$. We can regularize
this determinant 
as $R^{B_0}$, where
$B_0$ is the dimension of the space of 0-forms in the two 
dimensional world-sheet (regularized in a lattice, for instance).
With this prescription\footnote{This way of regularizing the
determinants reproduces the correct modular
anomaly under S-duality in abelian gauge theories,
as we will see in section 2.}
one realizes that the usual
measure in phase space:
${\cal D}\theta{\cal D}p_\theta$ gives upon integration on
$p_\theta$: ${\cal D}\theta R^{B_1}$, where $B_1$ is the 
dimension of 
the space of 1-forms in the world-sheet and emerges because
the momenta are 1-forms. Therefore it differs from our
definition of measure in configuration space. In order to
reproduce the partition function in configuration space we
need to include explicit factors of $R$ in the definition of
the measure in phase space. One can check that considering these
factors the correct shift of the dilaton is obtained after
performing the canonical transformation. These arguments 
however are only rigorous for constant backgrounds. We believe
that a similar reasoning could be applied to the general case.

\subsection{Non-abelian duality}

The conventional gauging approach to non-abelian duality 
\cite{DLOQ} has two
important drawbacks. The first one is that the transformation is
in general non-invertible (i.e. it is not possible to recover the
original theory by repeating the gauging procedure starting 
from the dual)
and the second that it is not valid
for non-spherical world-sheets \cite{AAGBL,GR}. 
A canonical transformation
description would be in this sense very useful, since it is 
invertible and the generalization to arbitrary
genus Riemann surfaces is straightforward, as we have seen. 
Such a description is in fact known for
those sigma models in which the non-abelian isometry acts without
isotropy, i.e. without fixed points. The most general sigma model
of this kind is \cite{GR}:
\bea
\label{2once}
S[g,x]=&&\int d\sigma_+d\sigma_-[E_{ab}(x)(\partial_+g 
g^{-1})^a (\partial_-g g^{-1})^b+F^R_{a\alpha}(x)
(\partial_+g g^{-1})^a
\partial_- x^\alpha+ \nonumber\\
&&F^L_{\alpha a}(x)\partial_+ x^\alpha 
(\partial_- g g^{-1})^a+F_{\alpha\beta}(x)\partial_+ x^\alpha
\partial_- x^\beta],
\eea
where $g\in G$, a Lie group (which we take to be compact), and
$\partial_\pm g g^{-1}=(\partial_\pm g g^{-1})^a T_a$ with 
$T_a$ the
generators of the corresponding Lie algebra\footnote{$\{T_a\}$
are normalized such that $Tr(T_aT_b)=\delta_{ab}$.}.
This model is invariant under right transformations
$g\rightarrow gh$, with $h\in G$.
Let us parametrize the
Lie group using the Maurer-Cartan forms $\Omega^a_k$, such that
\be
\label{2doce}
(\partial_\pm g g^{-1})^a=\Omega^a_k(\theta) \partial_\pm\theta^k.
\ee
 
The following canonical transformation from
$\{\theta^i,\Pi_i\}$ to $\{\chi^a,{\tilde \Pi}_a\}$:
\bea
\label{2trece}
&&\Pi_i=-(\Omega^a_i {\chi^\prime}^a+f_{abc}\chi^a
\Omega^b_j\Omega^c_i{\theta^\prime}^j) \nonumber\\
&&{\tilde \Pi}_a=-\Omega^a_i{\theta^\prime}^i
\eea
yields the non-abelian dual of (\ref{2once}) with respect to its
isometry $g\rightarrow gh$:
\be
\label{2catorce}
{\tilde S}=\int d\sigma_+ d\sigma_-
[(E+{\rm ad}\chi)^{-1}_{ab}(\partial_+\chi^a+F^L_{\alpha a}(x)
\partial_+ x^\alpha)(\partial_-\chi^b-F^R_{b\beta}(x)
\partial_- x^\beta)+F_{\alpha\beta}\partial_+x^\alpha
\partial_-x^\beta]
\ee
This was first realized in \cite{CZ1} for the case of 
$SU(2)$ principal
chiral models (where $E_{ab}=\delta_{ab}$, $F^R_{a\alpha}=
F^L_{\alpha a}=F_{\alpha\beta}=0$), generalized in 
\cite{L1,AL} to arbitrary group, and shown to apply also 
to this more general case in \cite{S}. 
(\ref{2trece}) reads, in configuration space variables:
\bea
\label{e1}
&&\Omega^a_i\partial_+\theta^i=-M_{ba}(\partial_+\chi^b+
F^L_{\alpha b}\partial_+ x^\alpha)=
-({\tilde g}_{ab}-{\tilde b}_{ab})\partial_+\chi^b-
({\tilde g}_{a\alpha}-{\tilde b}_{a\alpha})
\partial_+x^\alpha \nonumber\\
&&\Omega^a_i\partial_-\theta^i=M_{ab}(\partial_-\chi^b-
F^R_{b\alpha}\partial_-x^\alpha)=
({\tilde g}_{ab}+{\tilde b}_{ab})\partial_-\chi^b+
({\tilde g}_{a\alpha}+{\tilde b}_{a\alpha})
\partial_-x^\alpha.
\eea
These relations generalize (\ref{ab33}) to
non-abelian duality transformations, the main difference being 
that the components of the torsion in the Lie algebra variables
appear explicitly. 

(\ref{2trece}) is generated by:
\be
\label{2quince}
{\cal F}[\chi, \theta]=\oint d\sigma Tr(\chi\partial_\sigma
g g^{-1})
\ee
which is linear in the dual variables but non-linear in
the original ones. This means that in general it will receive
quantum corrections when implemented at the level of the Hilbert
spaces \cite{GC}, the reason being that we cannot prove a relation
like
\be
\label{e2}
|\chi^a\rangle=\int {\cal D}\theta^i(\sigma)
e^{i{\cal F}[\chi^a,\theta^i(\sigma)]}|\theta^i(\sigma)\rangle
\ee
using the eigenfunctions of the respective Hamiltonians.
However, it was shown in \cite{CZ1,L1} that such a relation
can in fact be proven using the eigenfunctions of the respective
conserved currents in the initial and dual theories. 
Of course
for this to be true we need to have a symmetry in the dual
theory, which is not the case for arbitrary backgrounds.
When $E_{[ab]}=F^L_{\alpha a}=F^R_{a\alpha}=0$ the original sigma
model is invariant under left transformations 
$g\rightarrow hg$, $h\in G$ and we also
find a symmetry in the dual theory under $\chi$ transforming
in the adjoint representation\footnote{This symmetry is the
reminiscence of the left symmetry of the original theory since
the left and right symmetries commute and we are dualizing with
respect to the right action only.}. Then it is easy to see that
the canonical transformation couples the conserved currents
associated to the left symmetry of the initial theory:
\be
\label{uu3}
J^{a(L)}_\pm=\frac12 E_{(ab)}\Omega^b_i\partial_\pm\theta^i
\ee
and the ones associated to $\chi\rightarrow h\chi h^{-1}$ in the
dual\footnote{Up to a total derivative term which for 
principal chiral models ($E_{ab}=\delta_{ab}$) is the responsible
for having curvature-free currents in the dual, that are coupled
to the curvature-free currents of the principal chiral model
\cite{CZ1,L1}.}:
\bea
\label{uu4}
&&{\tilde J}^a_+=\partial_+\chi^a-\frac12 E_{(ab)}M_{cb}
\partial_+\chi^c\nonumber\\
&&{\tilde J}^a_-=-\partial_-\chi^a+\frac12 E_{(ab)}M_{bc}
\partial_-\chi^c.
\eea
We can then show that
\be
\label{proc1}
{\tilde J}^a_\pm e^{i{\cal F}}=
J^{a(L)}_\pm e^{i{\cal F}}
\ee
and prove
(\ref{e2}) using the eigenfunctions of the respective
conserved currents.
{}From this relation 
the equivalence between the initial and dual theories
for arbitrary genus Riemann surfaces is straightforward
and it also allows the derivation of global
properties in the dual.
As in the abelian case there can still be renormalization effects
modifying the classical backgrounds.
We should also mention that a dilaton shift is needed in order
to preserve conformal invariance \cite{DLOQ}, exactly as in
the abelian case. This remains an open question within the
canonical transformation description whose resolution we
believe should be along the lines previously mentioned in the
abelian case.

\subsection{Superstrings}

The formulation of abelian \cite{H} and non-isotropic non-abelian 
\cite{CZ2,CUZ,S} T-dualities in $N=1$ superstring theories as  
canonical transformations is also known. 
Let us consider first the case of an abelian isometry
$\delta x^i=\epsilon k^i$, $i=1,\dots d$.
In adapted coordinates to the isometry 
$\{\theta,\psi_{\pm}^0,x^\alpha,\psi_{\pm}^\alpha\}$, 
$\alpha=1,\dots,d-1 ,$ 
we can write the $N=1$ action as:
\bea
\label{(1.4)}
&&S={1\over2}\int_{\Sigma}d^2\sigma\{g_{00}(\dot\theta^2 - 
\theta^{\prime 2}) + 2(\dot\theta + \theta^\prime)j_- + 
2(\dot\theta - \theta^\prime)j_+ 
-i (g_{00}\psi_{+}^0 + k_{\alpha}^{-}\psi_{+}^{\alpha})
(\dot\psi_{+}^0 - {\psi_{+}^0}^\prime)-\nonumber\\
&&i(g_{00}\psi_-^0+k_{\alpha}^+\psi_-^\alpha)(\dot\psi_{-}^0+
{\psi_-^0}^\prime)+V\}
\eea
where we have defined
\bea
\label{(1.5)}
&&j_{\pm} ={1\over2}(k_{\alpha}^{\mp}\partial_{\pm}x^\alpha + i
k_{[i,j]}^{\mp}\psi_{\pm}^j\psi_{\pm}^i) \nonumber\\
&&V=(g_{\alpha\beta}+b_{\alpha\beta})\partial_+ x^\alpha
\partial_- x^\beta-i\psi^i_+(g_{i\alpha}+b_{i\alpha})
\partial_-\psi_+^\alpha-i\psi^i_-(g_{i\alpha}-b_{i\alpha})
\partial_+\psi_-^\alpha- \nonumber\\
&&i\partial_j(g_{\alpha i}-b_{\alpha i})
\psi^i_+\partial_-x^\alpha\psi^j_+-i\partial_j
(g_{\alpha i}+b_{\alpha i})\psi^i_-\partial_+x^\alpha
\psi^j_-+\frac12 R^{(-)}_{ijkl}\psi^i_+\psi^j_+\psi^k_-\psi^l_-,
\eea
and $k_i^\pm=g_{0i}\pm b_{0i}$ as in the previous section.

The canonical momenta associated to the zero coordinates are
\bea
\label{(1.6)}
&&\Pi_{\pm}= 
{i\over2}(g_{00}\psi_{\pm}^0 + k_{\alpha}^{\mp}
\psi_{\pm}^\alpha)\\
\label{(1.61)} 
&&p_{\theta}= 
g_{00}\dot\theta + j_{+} + 
j_{-},
\eea
(\ref{(1.6)}) being two first class constraints.

The generating functional:
\be
\label{(1.8)}
{\cal F} = {1\over2}\oint d\sigma\{ \theta^\prime\tilde\theta - 
\theta\tilde\theta^\prime -
i\psi_{+}^0\tilde\psi_{+}^0 +i\psi_{-}^0\tilde\psi_{-}^0 \} 
\ee
induces the change of variables in phase space \cite{H}:
\bea
\label{(1.9)}
&&{\tilde\Pi}_{\pm} = -{\delta F\over\delta\tilde\psi_{\pm}^0} = 
\mp{i\over2}\psi_{\pm}^0, \qquad 
\Pi_{\pm} = {\delta F\over\delta\psi_{\pm}^0} = 
\mp{i\over2}\tilde\psi_{\pm}^0 \nonumber\\
&&p_{\tilde\theta} = -{\delta F\over\delta\tilde\theta} = - 
\theta^\prime ,
\qquad p_{\theta} 
= {\delta F\over\delta\theta} = - \tilde \theta^\prime,
\eea 
which yields the abelian T-dual with backgrounds given
by Buscher's formulas \cite{IKR}.
In configuration space this corresponds to:
\bea
\label{(1.9.1)}
&&{\psi}_\pm^0=\mp({\tilde g}_{00}{\tilde \psi}_\pm^0+
{\tilde k}_\alpha^\mp{\tilde \psi}_\pm^\alpha)\nonumber\\
&&\psi_\pm^\alpha={\tilde \psi}_\pm^\alpha
\eea
for the fermions, and:
\bea
\label{(1.9.2)}
&&\partial_+\theta=-{\tilde g}_{00}\partial_+{\tilde \theta}-
{\tilde k}^-_\alpha\partial_+{\tilde x}^\alpha-i
{\tilde k}^-_{[i,j]}{\tilde \psi}^j_+{\tilde \psi}^i_+=
-{\tilde k}^-_i\partial_+{\tilde x}^i-i{\tilde k}^-_{[i,j]}
{\tilde \psi}^j_+{\tilde \psi}^i_+
\nonumber\\
&&\partial_-\theta={\tilde g}_{00}\partial_-{\tilde \theta}+
{\tilde k}^+_\alpha\partial_-{\tilde x}^\alpha+i
{\tilde k}^+_{[i,j]}{\tilde \psi}^j_-{\tilde \psi}^i_- =
{\tilde k}^+_i\partial_-{\tilde x}^i+i{\tilde k}^+_{[i,j]}
{\tilde \psi}^j_-{\tilde \psi}^i_-,
\eea
for the bosons. (\ref{(1.9.1)}) and (\ref{(1.9.2)}) generalize
the abelian duality mapping (\ref{ab33}) to $N=1$ sigma models,
and can also be obtained from (\ref{ab33})
replacing bosonic fields by superfields and derivatives by
superderivatives. 

${\cal F}$ being linear in the 
original and dual variables implies that
the original and dual theories are also
equivalent quantum mechanically, as in the bosonic case.
As in that case its expression can be derived 
from the total time
derivative term that is induced in the gauging procedure.

Let us consider now non-isotropic non-abelian transformations
in $N=1$ sigma models.
For simplicity we
are going to restrict ourselves to the case of principal
chiral models: 
$g_{ij}=\Omega^a_i\Omega^a_j$ and $b_{ij}=0$.
Following \cite{CUZ} we can use tangent space variables 
for the fermions 
$\phi^a_\pm=\Omega^a_i \psi^i_\pm$ and consider an action:
\bea
\label{u1}
S&=&\int_\Sigma d\sigma_+ d\sigma_-[(\partial_+gg^{-1})^a
(\partial_-gg^{-1})^a-i\phi^a_+\partial_-\phi^a_+
-i\phi^a_-\partial_+\phi^a_-+\frac{i}{2}f_{abc}
\phi^a_+(\partial_-gg^{-1})^b\phi^c_++\nonumber\\
&&\frac{i}{2}f_{abc}\phi^a_-(\partial_+gg^{-1})^b\phi^c_-+
\frac18 f_{adb}f_{bce}\phi^a_+\phi^d_+\phi^c_-\phi^e_-].
\eea
Working in phase space variables $\{(\theta^i,\Pi_i),
(\phi^a_\pm,\Pi^a_{\phi\pm})\}$:
\bea
\label{u2}
&&\Pi_i=\Omega^a_i\Omega^a_j{\dot\theta}^j
+\frac{i}{4} f_{abc}
\Omega^b_i (\phi^a_+\phi^c_++\phi^a_-\phi^c_-)\\
\label{u3}
&&\Pi^a_{\phi\pm}=\frac{i}{2} \phi^a_\pm,
\eea
where
(\ref{u3}) are a set of first class constraints, the non-abelian
dual of (\ref{u1}) with respect to the right action of the whole
symmetry group $G$ can be obtained through a canonical 
transformation from 
$\{(\theta^i,\Pi_i)\}, (\phi^a_\pm,\Pi^a_{\phi\pm})\}$
to $\{(\chi^a,{\tilde \Pi}_a), ({\tilde \phi}^a_\pm, 
{\tilde \Pi}^a_{{\tilde \phi}\pm})\}$. In particular:
\bea
\label{u4}
&&\Pi_i=-(\Omega^a_i\chi^{\prime a}+f_{abc}\chi^a\Omega^b_j
\Omega^c_i\theta^{\prime j}) \nonumber\\
&&{\tilde \Pi}_a=-(\Omega^a_i \theta^{\prime i}+\frac{i}{4}
f_{abc}(\phi^b_+\phi^c_+-\phi^b_-\phi^c_-))
\eea
for the bosonic momenta, and:
\bea
\label{u5}
&&\Pi^a_{\phi\pm}=\mp\frac{i}{2}({\tilde \phi}^a_\pm+f_{abc}\chi^b
\phi^c_\pm) \nonumber\\
&&{\tilde \Pi}^a_{{\tilde \phi}\pm}=\mp \frac{i}{2}\phi^a_\pm
\eea
for the fermionic ones. 
Its generating functional is:
\be
\label{u6}
{\cal F}=\oint d\sigma [\chi^a \Omega^a_i\theta^{\prime i}
+\frac{i}{4}
f_{abc}\chi^a(\phi^b_+\phi^c_+-\phi^b_-\phi^c_-)-\frac{i}{2}
(\phi^a_+{\tilde \phi}^a_+-\phi^a_-{\tilde \phi}^a_-)].
\ee
The dual action is given by \cite{T}\footnote{In this reference
this dual action is derived following the gauging procedure.}:
\bea
\label{u7}
{\tilde S}&=&\int_{\Sigma}d\sigma_+ d\sigma_-[M_{ab}(\partial_+
\chi^a\partial_-\chi^b-i{\tilde \phi}^a_+\partial_-
{\tilde \phi}^b_++i\partial_+{\tilde \phi}^a_-
{\tilde \phi}^b_-)+ iM_{bc}f_{cde}M_{da}{\tilde \phi}^a_-
{\tilde \phi}^e_-\partial_+\chi^b+\nonumber\\
&&iM_{ac}f_{cde}M_{db}{\tilde \phi}^a_+{\tilde \phi}^e_+
\partial_-\chi^b+
L_{abcd}{\tilde \phi}^a_+{\tilde \phi}^b_-{\tilde \phi}^c_+
{\tilde \phi}^d_-], 
\eea
manifestly supersymmetric and where
$M=(1+{\rm ad}\chi)^{-1}$ and
$L_{abcd}=-(f_{agf}f_{ieb}+f_{aie}f_{gfb})M_{ci}M_{eg}M_{fd}$. 
The dual momenta are:
\bea
\label{u8}
&&{\tilde \Pi}_a=\frac12 (M_{(ab)}{\dot\chi}^b-M_{[ab]}
\chi^{\prime b}
-i(M{\rm ad}{\tilde \phi}_-M)_{ab}{\tilde \phi}^b_-
-i(M{\rm ad}{\tilde \phi}_+M)_{ba}{\tilde \phi}^b_+) \\
\label{u88}
&&{\tilde \Pi}^a_{{\tilde \phi}+}=\frac{i}{2}M_{ba}{\tilde \phi}^b_+
\nonumber\\
&&{\tilde \Pi}^a_{{\tilde \phi}-}=\frac{i}{2}M_{ab}{\tilde \phi}^b_-
\eea
where $({\rm ad}{\tilde \phi}_\pm)_{ab}=f_{abc}{\tilde \phi}^c_\pm$.
{}From (\ref{u88}) and (\ref{u5}) we see that the fermions simply 
transform with the change
of scale:
\bea
\label{u9}
&&\phi^a_+=-M_{ba}{\tilde \phi}^b_+ \nonumber\\
&&\phi^a_-=M_{ab}{\tilde \phi}^b_-.
\eea
The corresponding non-local transformation for the bosonic
part is given in terms of dual backgrounds by:
\bea
\label{mm2}
&&(\partial_+g g^{-1})^a=-({\tilde g}_{ab}-{\tilde b}_{ab})
\partial_+\chi^b-i\partial_e({\tilde g}_{ab}-
{\tilde b}_{ab}){\tilde \phi}^e_+{\tilde \phi}^b_+
-i(\phi_+^2)^a \nonumber\\
&&(\partial_-g g^{-1})^a=({\tilde g}_{ab}+{\tilde b}_{ab})
\partial_-\chi^b+i\partial_e({\tilde g}_{ab}+{\tilde b}_{ab})
{\tilde \phi}^e_-{\tilde \phi}^b_--i(\phi^2_-)^a,
\eea
where the last terms need still be written in terms of the
dual fermions. In this form we see that they generalize 
(\ref{(1.9.2)}) by means of the last quadratic terms in the
fermions, which are zero in the abelian case.
(\ref{u9}) and (\ref{mm2}) can also be obtained from 
the corresponding
(\ref{e1}) in superspace by introducing chiral
superfields \cite{CUZ}.
As in the pure bosonic non-abelian
case the canonical transformation couples the conserved currents
associated to the left symmetry $g\rightarrow hg$ of the initial
theory and the ones associated to transformations in the 
adjoint in the dual. Namely:
\be
\label{nn1}
J^{a(L)}_\pm=(\partial_\pm gg^{-1})^a+i(\phi^2_\pm)^a=
\Omega^a_i\partial_\pm\theta^i+\frac{i}{2}
f_{abc}\phi^b_\pm\phi^c_\pm 
\ee
with
\bea
\label{nn2}
{\tilde J}^a_+&=&\partial_+\chi^a-M_{ba}\partial_+\chi^b+
i(M{\rm ad}{\tilde \phi}_+M)_{ba}{\tilde \phi}^b_+
\nonumber\\
{\tilde J}^a_-&=&-\partial_-\chi^a+M_{ab}\partial_-\chi^b-
i(M{\rm ad}{\tilde \phi}_-M)_{ab}{\tilde \phi}^b_-.
\eea
{}From here we can also establish the quantum equivalence 
between the two theories. 

\subsection{Open Strings}
 
Recently there has been renewed interest in the study of 
open string theories
with the last developments in string dualities
(see for instance \cite{PCJ} and references therein).
In \cite{P} Polchinski showed that open strings with 
certain exotic boundary conditions (D-branes) were the 
carriers
of the RR charges required by string duality.
This identification allowed for many new tests of
string duality.
D-branes first arised as particular features under 
T-duality in
theories of open strings \cite{DLP,HG}. 
The duality transformation (\ref{1.1})
maps Neumann boundary conditions:
$\partial_n x=0$, to Dirichlet boundary conditions:
$\partial_t {\tilde x}=0$, where $\partial_n$ and $\partial_t$ 
are the normal and tangent derivatives to the boundary.
The ends of the strings are then confined to the 
${\tilde x}$ plane, which
is itself dynamical.
These particular objects with mixed Neumann and Dirichlet
boundary conditions are the D-branes \cite{DLP}.
For type I superstrings 
crosscap boundary conditions for the unoriented topologies are
mapped to orientifold conditions \cite{DLP,orien} and 
the dual D-brane
is hidden in the orientifold plane.
These dual theories may seem rather exotic, but they are just
a more suitable description at small distances
of the same original open string theory.

The open string-D-brane dualities of toroidal
compactifications have been
extended to more general backgrounds following the
gauging procedure to T-duality\footnote{In the first of
\cite{DO} there 
is also a brief
study with canonical transformations.}.
Namely, to
backgrounds with abelian \cite{ABB,DO} and
non-abelian isometries
\cite{FKS}. Certain backgrounds without isometries and more
recently WZW models have also
been studied in \cite{KS} and \cite{KS3} within the Poisson-Lie
T-duality.
We are going to discuss
T-duality for open strings in various
backgrounds within the canonical transformation approach
\cite{BL}. This approach is
particularly useful in obtaining
information about the boundary conditions,
since it provides an explicit mapping between initial and 
dual variables.

Let us consider open and closed strings propagating in a 
$d$ dimensional
background
of metric, antisymmetric tensor and abelian gauge field\footnote{We
will only consider abelian background
gauge fields. For a non-abelian treatment see the first of 
\cite{DO}.}.
In the neutral
case the action can be written:
\be
\label{1uno}
S=\int_\Sigma d\sigma_+ d\sigma_-(g_{ij}+b_{ij})
\partial_+ x^i \partial_- x^j+\int_{\partial\Sigma}
V_i \partial_t x^i 
\ee
where $V_i$ denotes the abelian background gauge field and
$\partial_t$ is the tangent derivative to the 
boundary\footnote{We consider $\sigma=$ constant boundaries
but in certain, specified, cases.}.
The boundary term
can be absorbed in the action by just considering:
\be
\label{1dos}
S=\int_\Sigma d\sigma_+ d\sigma_- (g_{ij}+B_{ij})
\partial_+ x^i \partial_- x^j
\ee
with $B_{ij}=b_{ij}+F_{ij}=b_{ij}+\partial_i V_j-\partial_j V_i$.
The torsion term is absent for the unoriented topologies.
Let us assume that there exists a Killing vector $k^i$ such that
${\cal L}_k g_{ij}=0$ and ${\cal L}_k B_{ij}=0$ (this means we can
have: ${\cal L}_k b_{ij}=\partial_i v_j-\partial_j v_i$ and
${\cal L}_k V_i=-v_i+\partial_i\varphi$, for some $v_i$, 
$\varphi$). The dual with respect to this isometry can be
constructed as in the closed string case by performing the
canonical transformation:
\bea
\label{op1}
&&p_\theta=-{\tilde \theta}^\prime\nonumber\\
&&p_{\tilde \theta}=-\theta^\prime
\eea
the only difference being that $b$ is replaced by $B$ in order
to absorb the background gauge field. 
In this case we also need to care about the boundary conditions.
The canonical transformation approach is particularly adequate 
to deal with boundary conditions since it provides the
explicit relation between the target space
coordinates of the original and dual theories. 
{}From (\ref{op1})
we get\footnote{We use capital letters for the dual backgrounds
to account for its dependence on the abelian background gauge
field.}:
\be
\label{1nueve}
{\dot {\tilde \theta}}=-(g_{00}\theta^\prime +g_{0\alpha}
{\theta^\prime}^\alpha -B_{0\alpha} {\dot x}^\alpha)
\ee
and
\be
\label{1diez}
g_{0\alpha}\theta^\prime+g_{\alpha\beta}x^{\prime\beta}+
B_{0\alpha}{\dot \theta}-B_{\alpha\beta}{\dot x}^\beta= 
{\tilde G}_{0\alpha}{\tilde \theta}^\prime+{\tilde G}_{\alpha\beta}
x^{\prime\beta}+{\tilde B}_{0\alpha}{\dot {\tilde \theta}}-
{\tilde B}_{\alpha\beta}{\dot x}^\beta .
\ee
Then, Neumann boundary conditions for the original theory 
\cite{CLNY}:
\be
\label{1siete}
g_{ij}x^{\prime j}-B_{ij}{\dot x}^j=0
\ee
imply in the dual:
\bea
\label{1once}
&&{\dot {\tilde \theta}}=0 \nonumber\\
&&{\tilde G}_{0\alpha}{\tilde \theta}^\prime
+{\tilde G}_{\alpha\beta}
x^{\prime\beta}+{\tilde B}_{0\alpha}{\dot {\tilde \theta}}-
{\tilde B}_{\alpha\beta}{\dot x}^\beta=0.
\eea
These mixed boundary conditions represent a flat Dirichlet 
$(d-2)$-brane in
the dual background \cite{ABB}. Also
from here we can deduce the collective
motion of the brane. Decomposing 
$B_{0\alpha}=b_{0\alpha}-\partial_\alpha V_0$ we realize that the
usual Buscher's backgrounds for closed strings (with the torsion
$b$) are gotten provided we redefine 
${\tilde \theta}\equiv {\tilde \theta}+V_0(x^\alpha)$.
Therefore $V_0(x^\alpha)$ gives the transverse position of the
brane in the dual theory\footnote{A particular case is when
$V_0$ is taken pure gauge locally breaking $U(N)$ to $U(1)^N$,
i.e. when a Wilson line 
$V_0={\rm diag}\{\theta_1,\dots,\theta_N\}$ is included.
In this case we get a maximum of $N$ D-branes in the dual
theory with fixed positions at $\theta_i$, $i=1,\dots,N$
\cite{PCJ}.}.
If we dualize $n$ commuting 
isometries it is straightforward to check that a  Dirichlet 
$(d-n-1)$-brane is obtained in the dual. 
It is perhaps worth mentioning that there are some particular 
backgrounds (those whose conserved currents associated to the 
isometry are chiral
\cite{RV}) which are at the same time backgrounds of open 
strings and D-branes depending on the boundary conditions, 
which are in turn
related by a
T-duality transformation.

Let us now analyze the unoriented topologies.
Invariance under world-sheet parity implies that the
antisymmetric tensor and the abelian gauge field are projected
out of the spectrum. We can still have non-abelian gauge fields
in $SO(N)$ and $USp(N)$ but they must be treated differently
(see for instance the first in \cite{DO}).
Unoriented topologies
can be obtained from oriented ones by identifications
of points on the boundary \cite{GSW}. For instance the projective
plane is obtained from the disk\footnote{We have to make first a
Wick rotation to imaginary time.}
identifying opposite points. The 
topology thus obtained is a crosscap.  
Under abelian T-duality we should get the mapping from crosscap
to orientifold conditions \cite{ABB}.
Crosscap boundary conditions for the coordinate 
adapted to the isometry:
\be
\label{1doce}
{\dot \theta}(\sigma+\pi)=-{\dot \theta}(\sigma)\qquad
\theta^\prime(\sigma+\pi)=\theta^\prime(\sigma),
\ee
where we are parametrizing the boundary of the disk by
$(0,2\pi)$ and identifying opposite points:
$\theta(\sigma+\pi)=\theta(\sigma)$,
translate to:
\be
\label{1trece}
p_\theta(\sigma+\pi)=-p_\theta(\sigma)\qquad
\theta^\prime (\sigma+\pi)=\theta^\prime (\sigma)
\ee
in phase space. Then
(\ref{op1}) implies:
\be
\label{1catorce}
{\tilde \theta}^\prime (\sigma+\pi)=
-{\tilde \theta}^\prime (\sigma)\qquad
p_{{\tilde \theta}}(\sigma+\pi)=p_{{\tilde \theta}}(\sigma),
\ee
which are orientifold conditions in phase space since
$p_{{\tilde \theta}}(\sigma+\pi)=p_{{\tilde \theta}}(\sigma)$
implies ${\dot {\tilde \theta}}(\sigma+\pi)={\dot 
{\tilde \theta}}(\sigma)$\footnote{Integration on the first 
equation implies ${\tilde \theta}(\sigma+\pi)=
-{\tilde \theta}(\sigma)$ so that the world-sheet 
parity reversal is accompanied by a $Z_2$ transformation in
space-time. These kinds of constructions are the orientifolds
\cite{DLP,orien}.}.
The orientifold plane is at ${\tilde \theta}=0$ and it's
non-dynamical, because the abelian
gauge field is zero in the unoriented case.
The rest of the coordinates still
satisfy crosscap boundary conditions. 

We can make an analogous analysis for the non-abelian 
backgrounds (\ref{2once}) \cite{BL}.
The abelian gauge fields that are compatible
with the non-abelian isometry $g\rightarrow gh$ have the
form $V_i=\frac12 \Omega^a_i C^a(x)$, with $C$ arbitrary,
and $V_\alpha$ $\theta^i$-independent. 
Then $E_{[ab]}=b_{ab}+f_{abc}C^c(x)$, where $b_{ab}$ is
the closed strings antisymmetric tensor and
$F^R_{a\alpha}=f^R_{a\alpha}-\frac12 \partial_\alpha
C^a(x)$, $F^L_{\alpha a}=f^L_{\alpha a}+\frac12
\partial_\alpha C^a(x)$ (with $f^R, f^L$ the corresponding
closed strings backgrounds).
The canonical transformation (\ref{2trece}) implies that the 
dual of Neumann boundary conditions are mixed Dirichlet and
Neumann conditions weighted with the initial metric and torsion
respectively, for the coordinates transforming
under the isometry group. The rest of the coordinates still 
satisfy Neumann
boundary conditions in the dual. 
However in those cases in which we can establish a
correspondence between the Hilbert spaces of the initial and 
dual theories (when the dual theory admits an isometry) the dual
boundary conditions for the coordinates transforming under the
isometry reduce to generalized Dirichlet conditions:
\be
\label{op2}
{\tilde g}_{ab}\dot\chi^b-{\tilde b}_{ab}\chi^{\prime b}=0,
\ee
generalized in the sense that it is the momentum associated to a
non-flat background which vanishes at
the ends of the string. 
We can then conclude that for 
certain kinds
of sigma models with non-abelian isometries a curved 
$(d-{\rm dim}G-1)$ D-brane with metric
${\tilde g}_{ab}$ and torsion ${\tilde b}_{ab}$ is obtained
in the dual. The backgrounds of
unoriented strings are among the ones for which we get Dirichlet
boundary conditions. In these cases we can also study the 
mapping of crosscap boundary conditions. The result is that in the
dual, generalized orientifold conditions:
\bea
\label{2veinticuatro}
&&{\tilde g}_{ab}\chi^{\prime b}(\sigma+\pi)
-{\tilde b}_{ab}\dot{\chi}^b(\sigma+\pi)=-
({\tilde g}_{ab}\chi^{\prime b}(\sigma)-{\tilde b}_{ab}
\dot{\chi}^b(\sigma)) \nonumber\\
&&{\tilde g}_{ab}\dot{\chi}^b(\sigma+\pi)
-{\tilde b}_{ab}\chi^{\prime b}(\sigma+\pi)=
{\tilde g}_{ab}\dot{\chi}^b(\sigma)-{\tilde b}_{ab}
\chi^{\prime b}(\sigma)
\eea
are satisfied, meaning that the momentum must be equal at the
identification points but the momentum flows out of them must
have opposite sign.

\subsubsection{Superstrings}

In a similar way we can study the mapping of the boundary conditions
in open superstring theory \cite{BL}.
We consider abelian background gauge fields, which are absorbed in
a torsion term $B_{ij}=b_{ij}+F_{ij}$.
Starting with the usual Neumann R-NS
boundary conditions\footnote{R-NS are minima of the bulk action
and therefore are classical boundary conditions only if
$B_{ij}=0$. However the simplest minima of the full action give
trivial dynamics for the fermions at the boundary, being in this
sense too restrictive. One can study if this is also the case for
arbitrary minima and find that in general the fermionic
contribution at the boundary is not zero \cite{BL}. Here we
are still going to consider R-NS for simplicity.}
for the initial theory:
\bea
\label{(1.22)}
&&g_{ij}x^{\prime j}-B_{ij}\dot{x}^j=0\nonumber\\
&&\psi^i_+=\eta\psi^i_-;\qquad \eta=\pm 1
\eea
we get after an abelian duality transformation
(\ref{(1.9.1)}), (\ref{(1.9.2)}): 
\bea
\label{(1.23)}
&&{\tilde\psi}^\alpha_+=\eta{\tilde\psi}^\alpha_-\nonumber\\
&&{\tilde\psi}^0_++\eta{\tilde\psi}^0_-=2\eta k_\alpha^*
{\tilde\psi}^\alpha_-\nonumber\\
&&\dot{\tilde\theta}=i\partial_\alpha k_\beta^*
{\tilde\psi}^\alpha_-
{\tilde\psi}^\beta_-\nonumber\\
&&{\tilde G}_{\alpha i}{\tilde x}^{\prime i}-{\tilde B}_{\alpha i}
{\dot{\tilde x}}^i=-ik_\alpha^*\partial_\beta{\tilde k}^2
{\tilde\psi}^\beta_-
{\tilde\psi}^0_-+i({\tilde k}^\alpha_+\partial_\beta
k^*_\sigma-k^*_\alpha
\partial_\beta
{\tilde k}^+_\sigma){\tilde\psi}^\beta_-{\tilde\psi}^\sigma_-
\eea
where $k^*_\alpha=B_{0\alpha}$. These results are in agreement 
with (3.12) in
\cite{ABB}.
The non-trivial terms (those that spoil Dirichlet NS-R boundary
conditions in the dual) are all proportional to $B_{0\alpha}$.
Therefore a super D-brane is obtained in the dual only if the
original background is such that $B_{0\alpha}=0$.
If this occurs
(\ref{(1.23)}) turns into: 
${\tilde\psi}^0_+=-\eta{\tilde\psi}^0_-$, accounting for the
reversal of space-time chirality under T-duality
\cite{DHS},
$\dot{\tilde\theta}=0$, and Neumann R-NS boundary 
conditions for the rest of the coordinates. 
This is the case, in particular, for the type I superstring
where the D-brane is actually an orientifold.
In this theory consistency conditions restrict the possible
D-manifolds to one, five and nine-branes \cite{PCJ}.
Since the only consistent open superstring theory is the
type I superstring, which contains unoriented topologies, it
is interesting to analyze in some more detail the
unoriented world-sheets. As in
the previous section we consider the projective plane,
obtained from the disk by identifying opposite points.
Crosscap boundary conditions for the fermions contain 
an $i$ factor
due to the fact that we are taking a constant time boundary
\cite{CLNY}:
\bea
\label{cr1}
&&\psi_{+}^i (\sigma + \pi) = i\eta \p^i (\sigma),\qquad 
\eta=\pm 1 \\
\label{cr2}
&&x^{\prime i}(\sigma + \pi) = x^{\prime i}(\sigma),\qquad
\dot x^i(\sigma + \pi) = -\dot x^i(\sigma).
\eea
These conditions 
are mapped under (\ref{(1.9.1)}) and (\ref{(1.9.2)}) into:
\be
\label{cr3}
\tilde \psi_{+}^\alpha (\sigma + \pi) = 
i\eta\tilde \p^\alpha (\sigma) \qquad
\tilde\psi_{+}^0 (\sigma + \pi) = -i\eta \tilde\p^0 (\sigma)
\ee
for the fermions, giving the usual change of sector for the
0-component, and to:
\bea
\label{(1.28)}
&&\dot{\tilde\theta}(\sigma + \pi) = \dot{\tilde\theta}(\sigma)
\qquad \dot{{\tilde x}}^\alpha(\sigma+\pi)
=-\dot{{\tilde x}}^\alpha(\sigma)\nonumber\\
&&\tilde{\theta^\prime}(\sigma + \pi) = 
-\tilde{\theta^\prime}(\sigma) \qquad
{\tilde x}^{\prime\alpha}(\sigma+\pi)=
{\tilde x}^{\prime\alpha}(\sigma)
\eea
for the bosons, i.e. orientifold conditions for the 
${\tilde \theta}$ coordinate and crosscap for the rest. 
Therefore the dual theory is an orientifold, static since the 
abelian electromagnetic field is absent for unoriented strings.

Let us now concentrate on the non-abelian models (\ref{u1})
\cite{BL}. 
Restricting to the
case of R-NS boundary conditions for the fermions we get 
in the dual:
\be
\label{u11}
{\tilde \phi}^a_+=-\eta M_{ba}^{-1}M_{bc}{\tilde \phi}^c_-,
\ee
which are not NS-R boundary conditions. We can just point out
that they could be interpreted as NS-R plus corrections in
${\rm ad}\chi$.

Concerning the bosons, if we start with Neumann boundary 
conditions:
$\Omega^a_i\theta^{\prime i}=0$, we obtain in terms of the
dual backgrounds:
\be
\label{u13}
{\tilde g}_{ab}\dot{\chi}^b-{\tilde b}_{ab}\chi^{\prime b}+
\frac{i}{2}\partial_e({\tilde g}_{ab}+{\tilde b}_{ab})
{\tilde \phi}^e_-{\tilde \phi}^b_-+\frac{i}{2}\partial_e
({\tilde g}_{ab}-{\tilde b}_{ab}){\tilde \phi}^e_+
{\tilde \phi}^b_+=0.
\ee
This equation represents the vanishing of ${\tilde \Pi}_a$
(given by (\ref{u8}))
at the ends of the string. Therefore we find a curved
$N=1$ supersymmetric D-brane
(in this particular example (-1)-brane, since we haven't 
allowed
for inert coordinates) with metric and torsion given by
${\tilde g}_{ab}$ and ${\tilde b}_{ab}$.
However since the only consistent open superstring theory 
contains unoriented topologies the D-brane is an orientifold,
as happened in the abelian case. In particular, one can see
that crosscap boundary conditions are mapped to:
\bea
\label{z2}
&&{\tilde \Pi}_a(\sigma+\pi)={\tilde \Pi}_a(\sigma)\nonumber\\
&&({\tilde g}_{ab}\chi^{\prime b}-{\tilde b}_{ab}\dot{\chi}^b
+\frac{i}{2}\partial_e({\tilde g}_{ab}-{\tilde b}_{ab})
{\tilde \phi}^e_+{\tilde \phi}^b_+-\frac{i}{2}\partial_e
({\tilde g}_{ab}+{\tilde b}_{ab}){\tilde \phi}^e_-
{\tilde \phi}^b_-)|_{\sigma+\pi}=\nonumber\\
&&-({\tilde g}_{ab}\chi^{\prime b}-{\tilde b}_{ab}
\dot{\chi}^b+\frac{i}{2}\partial_e({\tilde g}_{ab}-
{\tilde b}_{ab}){\tilde \phi}^e_+{\tilde \phi}^b_+-
\frac{i}{2}\partial_e({\tilde g}_{ab}+{\tilde b}_{ab})
{\tilde \phi}^e_-{\tilde \phi}^b_-)|_\sigma,
\eea
where the second equation represents that the 
momenta flowing out
of the identification points must have opposite signs,
as in the bosonic
non-abelian case.
The dual fermions satisfy:
\be
\label{z3}
{\tilde \phi}^a_+(\sigma+\pi)=
-i\eta M^{-1}_{ba}M_{bc}{\tilde \phi}^c_-(\sigma).
\ee

\section{S-duality in Gauge Theories}
\setcounter{equation}{0}

\subsection{Abelian gauge theories}

Four dimensional
abelian gauge theories are invariant under strong-weak coupling 
duality\footnote{For reviews on S-duality in gauge theories
see for instance \cite{OHDV}.},
generated by the interchange of the electric and magnetic
degrees of freedom of the theory: 
$dA\rightarrow *dA$, with $A$ the abelian gauge field.
Both T-duality in two dimensional sigma 
models\footnote{What follows holds for
toroidal compactifications. We have seen in the previous sections
the generalization to other backgrounds.} and S-duality in four
dimensional
abelian gauge theories are generated by the mapping
$d\rightarrow *d$ in the corresponding two dimensional 
world-sheet or
four dimensional space-time.
In fact, they are particular cases of a more
general duality present in $d$ dimensional theories of
$p$ forms \cite{barbon,QT}:
\be
\label{proc3}
S\sim \int \frac{1}{g^2}d^d x\, dA_p\wedge *dA_p
\ee
(in 2 dim $g$ is the inverse of the compactification radius)
where the mapping $d\rightarrow *d$ yields a dual
theory formulated in terms of $(d-p-2)$ forms. 
This duality can also be described as a canonical transformation
in the corresponding phase space, as one would expect.
In this section we study in detail the canonical transformation
description of S-duality in Maxwell theory ($d=4, p=1$) 
\cite{L2}. The generalization
to arbitrary $d,p$ will also be described at the end. 

Let us consider the Maxwell Lagrangian with $\theta$-term:
\bea
\label{bru1}
L&=&\frac{1}{8\pi}(\frac{4\pi}{g^2}F_{mn}F^{mn}+
\frac{i\theta}{4\pi}
\epsilon_{mnpq}F^{mn}F^{pq}) \nonumber\\
&=&\frac{i}{8\pi}({\bar \tau}
F^+_{mn}F^{+mn}
-\tau F^-_{mn}F^{-mn})
\eea
defined on a Euclidean four-manifold $M_4$.
Here $\tau=\theta/2\pi+4\pi i/g^2$,
$F_{mn}=\partial_m A_n-\partial_n A_m$,
$\,^*F_{mn}=\frac12 \epsilon_{mnpq}F^{pq}$
and
$F^\pm_{mn}=\frac12 (F_{mn}\pm\,^*F_{mn})$.

The canonical momenta are given by\footnote{We have dropped the
global $i/8\pi$ factor. It will then appear when exponentiating
these quantities.}:
\bea
\label{bru2}
&&\Pi_0=0\nonumber\\
&&\Pi^\alpha=
4{\bar \tau} F^{+0\alpha}-4\tau F^{-0\alpha},
\eea
where $\alpha$ runs over spatial indices, and the
Hamiltonian:
\be
\label{bru3}
H=\frac{1}{4(\bar{\tau}-\tau)}\Pi_\alpha \Pi^\alpha+
\partial_\alpha A_0
\Pi^\alpha-\frac{\bar{\tau}+\tau}{\bar{\tau}-\tau}\Pi_\alpha
\,^*F^{0\alpha}
+\frac{4\bar{\tau}\tau}{\bar{\tau}-\tau}\,^*F^{0\alpha}
\,^*F_{0\alpha}.
\ee
The primary constraint $\Pi_0=0$ implies the secondary
$\partial_\alpha\Pi^\alpha=0$, therefore we can drop the
$\partial_\alpha A_0\Pi^\alpha$ term keeping in mind that
the Hamiltonian is defined in this restricted phase space.

The interchange between electric and magnetic degrees of 
freedom can be written as the following canonical transformation in
the phase space of the theory\footnote{Note that in the definition
of $\Pi^\alpha$ there is also a contribution from $\,^*F^{0\alpha}$
when $\theta\ne 0$ \cite{Wi}.}:
\bea
\label{bru4}
&&{\Pi}^\alpha=
-4\,^*{\tilde F}^{0\alpha}\nonumber\\
&&{\tilde \Pi}^\alpha
=4\,^*F^{0\alpha},
\eea
where ${\tilde F}=d{\tilde A}$, with generating functional:
\be
\label{bru5}
{\cal F}=-2\int_{M_3} d^3 x ({\tilde A}_\alpha\,^*F^{0\alpha}+
A_\alpha\,^*{\tilde F}^{0\alpha})=-\int_{D_4/\partial D_4=M_3} d^4 x
dA\wedge d{\tilde A}.
\ee
(\ref{bru4})
yields the following Hamiltonian:
\be
\label{bru6}
{\tilde H}=\frac14 \frac{\bar{\tau}\tau}{\bar{\tau}-\tau}
{\tilde \Pi}_\alpha
{\tilde \Pi}^\alpha+
\frac{\bar\tau
+\tau}{\bar{\tau}-\tau}{\tilde \Pi}_\alpha\,^*{\tilde F}^{0\alpha}+
\frac{4}{\bar{\tau}-\tau}\,^*{\tilde F}_{0\alpha}
\,^*{\tilde F}^{0\alpha},
\ee
in which ${\tilde \tau}=-1/\tau$.
Since the original Hamiltonian is defined in the restricted
phase space given by
$\Pi_0=0$, $\partial_\alpha\Pi^\alpha=0$
 we need to analyse as
well the mapping of the constraints and check that the dual
Hamiltonian is defined in the same restricted phase space.
The constraint ${\tilde \Pi}_0=0$ is straightforwardly
obtained from the generating functional, since there is no
dependence on ${\tilde A}_0$. The secondary constraint 
$\partial_\alpha {\tilde \Pi}^\alpha=0$
is obtained from the Bianchi identity 
$\partial_\alpha\,^*F^{0\alpha}$ of the original theory
and finally the constraint $\partial_\alpha\Pi^\alpha=0$
implies that ${\tilde F}$ is derived from a vector potential
${\tilde A}$ as a consequence of Poincar\`e's lemma.
We can introduce a $\partial_\alpha{\tilde A}_0
{\tilde \Pi}^\alpha$ term in the Hamiltonian imposing
the constraint $\partial_\alpha{\tilde \Pi}^\alpha=0$
as in (\ref{bru3}) and finally read a dual Lagrangian:
\be
\label{bru7}
{\tilde L}=\frac{i}{8\pi}(-\frac{1}{{\bar \tau}} 
{\tilde F}^+_{mn} {\tilde
  F}^{+mn}
+\frac{1}{\tau} {\tilde F}^-_{mn}{\tilde F}^{-mn}).
\ee
Some useful information can be obtained within this approach.
The generating functional is linear in both the original
and dual variables. We can then write:
\be
\label{bru9}
\psi_k[{\tilde A}]=N(k)\int {\cal D}A(x^\alpha) 
e^{\frac{i}{8\pi}{\cal F}[{\tilde
    A},A(x^\alpha)]}
\phi_k[A(x^\alpha)]
\ee
with $\phi_k[A]$ and $\psi_k[{\tilde A}]$ eigenfunctions of the
initial
and dual Hamiltonians respectively.
{}From this relation global properties can be easily worked out.
The Dirac quantization condition:
\be
\label{bru10}
\int_{\Sigma}F=2\pi n,\quad n\in Z,
\ee
for $\Sigma$ homologically non-trivial two-cycles in the manifold,
implies the same quantization in the dual:
\be
\label{bru11}
\int_{\Sigma}{\tilde F}=2\pi m,\quad m\in Z.
\ee
We can also analyze the transformation of the partition function in
phase space\footnote{The integration over $\Pi_0$ is canceled by
the gauge group volume and integration on $A_0$ yields the
constraint $\partial_\alpha\Pi^\alpha=0$, which in the dual theory
implies that ${\tilde F}$ is derived from a vector potential.}:
\be
\label{bru12}
Z_{ps}=\int {\cal D}A_m {\cal D}\Pi^\alpha e^{-
\frac{i}{8\pi}\int d^4x (\dot{A}_\alpha
\Pi^\alpha-H)}.
\ee
Under (\ref{bru4})
${\cal D}A_\alpha {\cal D}\Pi^\alpha={\cal D}
{\tilde A}_\alpha {\cal D}
{\tilde \Pi}^\alpha$
and therefore ${\tilde Z}_{ps}=Z_{ps}$ (we have previously seen
that both theories are defined in the same restricted phase
space),
which implies that in phase space the partition function is 
invariant under duality. 
However we know that it should transform as a 
modular form with a given modular weight \cite{witten}.
This modular factor appears when going to configuration space,
as we are going to show. 

Integrating the momenta in (\ref{bru12}) gives:
\be
\label{bru13}
Z_{ps}=\int {\cal D}A_m ({\rm Im} \tau)^{B_2/2} e^{-\int d^4x L}
\ee
with $L$ given by (\ref{bru1}). 
The factor $({\rm Im} \tau)^{B_2/2}$ in the
measure is the regularized $({\rm det\,Im}\tau)^{1/2}$ coming 
from the gaussian integration over the momenta. 
$B_2$ is the dimension of the
space of 2-forms in the four dimensional manifold $M_4$ 
(regularized on a lattice) and emerges because the momenta 
are 2-forms. The same calculation in the dual phase space 
partition function gives:
\be
\label{bru14}
{\tilde Z}_{ps}=\int {\cal D}{\tilde A}_m ({\rm Im}\tau)^{B_2/2}
{\bar \tau}^{-B_2^+/2}\tau^{-B_2^-}
e^{-\int d^4x {\tilde L}}
\ee
with ${\tilde L}$ given by (\ref{bru7}), where we have regularized
\be
\label{regu1}
({\rm det}({\rm Im} -\frac{1}{\tau}))^{1/2}=
({\rm Im} \tau)^{B_2/2}\bar\tau^{-B_2^+/2}\tau^{-B_2^-/2}
\ee
and $B_2^+$ ($B_2^-$) is the dimension of the
space of self-dual
(anti-self-dual) 2-forms.
In configuration space the partition function is defined 
by \cite{witten}:
\be
\label{bru15}
Z=({\rm Im} \tau)^{(B_1-B_0)/2}\int {\cal D}A_m e^{-S}=
({\rm Im} \tau)^{(B_1-B_0-B_2)/2} Z_{ps}
\ee
and in the dual model
\be
\label{bru16}
{\tilde Z}=(\frac{{\rm Im} \tau}{\tau \bar\tau})^{(B_1-B_0)/2}
\int {\cal D}
{\tilde A}_m e^{-{\tilde S}}=({\rm Im}\tau)^{(B_1-B_0-B_2)/2}
\tau^{(\chi-\sigma)/4}{\bar \tau}^{(\chi+\sigma)/4}
{\tilde Z}_{ps},
\ee
where $\chi=2(B_0-B_1)+B_2$ is the Euler number (the 
regularization is such
that $B_p=B_{4-p}$) and $\sigma=B_2^+-B_2^-$ is
the signature of the manifold.
{}From $Z_{ps}={\tilde Z}_{ps}$ we get
\be
\label{bru17}
Z=\tau^{-(\chi-\sigma)/4}{\bar\tau}^{-(\chi+\sigma)/4} 
{\tilde Z}.
\ee
Therefore in configuration space
the partition function transforms as a modular form
\cite{witten}.
It is clear that the solution to this puzzle is that the
regularization prescription for the determinants is such
that $Z$ is not obtained from $Z_{ps}$ after integrating out
the momenta, as it is clearly shown in (\ref{bru15})
and (\ref{bru16}). 
If we impose this requirement to the phase space partition
function the corresponding factors in the measure need to
be introduced and we also obtain that it transforms as 
(\ref{bru17}).

The same analysis made for Maxwell's theory can be 
straightforwardly generalized to $p$-forms
abelian gauge theories in $d$ dimensions \cite{L2}.
In this case the generating functional and
corresponding canonical transformation are:
\be
\label{3dos}
{\cal F}=-\int_{D_d/\partial D_d=M_{d-1}} d^dx 
dA\wedge d{\tilde A}.
\ee
\bea
\label{3tres}
&&\Pi^{\alpha_1\ldots\alpha_p}=\frac{\delta {\cal F}}{\delta
A_{\alpha_1\ldots\alpha_p}}=
-(p+1)!(d-p-1)!\,^*{\tilde F}^{0\alpha_1\ldots\alpha_p}\nonumber\\
&&{\tilde \Pi}^{\alpha_1\ldots\alpha_{d-p-2}}=-
\frac{\delta F}{\delta{\tilde A}_{\alpha_1\ldots\alpha_{d-p-2}}}=
(p+1)!(d-p-1)!\,^*F^{0\alpha_1\ldots\alpha_{d-p-2}}.
\eea
The phase space partition function is invariant under these
transformations. However upon integration on the momenta we
get for the partition function in configuration space 
\cite{barbon}: $Z=(4\pi/g^2)^{(-1)^p\chi/2}{\tilde Z}$ 
in general, and 
$Z=\tau^{-(\chi-\sigma)/4}{\bar \tau}^{-(\chi+\sigma)/4}{\tilde Z}$
for $d=2(p+1)$, $p$ odd, i.e. when a $\theta$-term is allowed in
the theory.

\subsection{Non-abelian gauge theories}

Non-supersymmetric Yang-Mills theories are not
invariant under the interchange of the electric and magnetic
degrees of freedom. However following a first order formalism
or a gauging-type procedure \cite{halpern}
it is possible to construct the
dual of the theory, which turns out to be a Freedman-Townsend's 
type of theory \cite{ft},
depending on 2-forms that are not derivable from
a vector potential. A generalized duality transformation
relating it to the initial Yang-Mills theory has been given
in \cite{HFS} in terms of loop variables.

It is also possible to construct the dual theory by performing
a simple canonical transformation in phase space, as we are
now going to show. Written in configuration space variables
this transformation provides the generalization to the non-abelian
case of the 
electric-magnetic mapping of abelian gauge theories. 

Starting with the Yang-Mills Lagrangian for a compact group $G$
on a Euclidean manifold:
\bea
\label{4uno}
L&=&\frac{1}{8\pi}(\frac{4\pi}{g^2}F^{(a)}_{mn} F^{(a)mn}+
\frac{i\theta}{4\pi}\epsilon^{mnpq}F^{(a)}_{mn} F^{(a)}_{pq})
\nonumber\\
&=&\frac{i}{8\pi}({\bar \tau} F^{(a)+}_{mn} F^{(a)+ mn}-\tau
F^{(a)-}_{mn}F^{(a)- mn})
\eea
where $F=dA-A\wedge A$ and we have chosen $Tr(T^a T^b)=\delta^{ab}$
($T^a$ are the generators of the Lie algebra) we can construct
the Hamiltonian:
\be
\label{i2}
H=\frac14 \frac{1}{{\bar \tau}-\tau}\Pi^a_\alpha \Pi^{a\alpha}+
(\partial_\alpha A^a_0+f_{abc}A^b_0A^c_\alpha)\Pi^{a\alpha}-
\frac{{\bar \tau}+\tau}{{\bar \tau}-\tau}\Pi^{a\alpha}\,
^*F^{(a)}_{0\alpha}+\frac{4{\bar \tau}\tau}{{\bar \tau}-\tau}
\,^*F^{(a)}_{0\alpha}\,^*F^{(a)0\alpha},
\ee
where $f_{abc}$ are the structure constants of the Lie algebra
and
\bea
\label{i1}
&&\Pi^{a\alpha}=
2({\bar \tau}-\tau)F^{(a)0\alpha}+2({\bar \tau}+\tau)\,
^*F^{(a)0\alpha} \nonumber\\
&&\Pi^{a0}=0.
\eea
The interchange between electric and magnetic degrees of
freedom:
\bea
\label{41bc}
&&\Pi^{a\alpha}=-4\,^*{\tilde F}^{(a)0\alpha} \nonumber\\
&&{\tilde \Pi}^{a\alpha}=4\,^*F^{(a)0\alpha},
\eea
where ${\tilde F}=d{\tilde A}-{\tilde A}\wedge {\tilde A}$,
yields a Hamiltonian in which the coupling has transformed
as $\tau\rightarrow -1/\tau$. However in the non-abelian
case (\ref{41bc}) is not a canonical transformation since
the Poisson brackets are not left invariant. Moreover the
secondary constraints:
\be
\label{41bf}
\partial_\alpha \Pi^{a\alpha}-f_{abc}A^b_\alpha\Pi^{c\alpha}=0
\ee
of Yang-Mills, imply in the dual theory:
\be
\label{cor8}
\partial_\alpha\,^*{\tilde F}^{(a)0\alpha}-
f_{abc}A^b_\alpha ({\tilde F})
\,^*{\tilde F}^{(c)0\alpha}=0
\ee
from where
in the absence of a non-abelian analogue of Poincar\`e's
lemma we cannot conclude that ${\tilde F}$ is derived from
a vector potential.

In fact
we know from previous calculations \cite{halpern} that the dual
theory is given by:
\be
\label{cor9}
{\tilde L}=\frac{i}{8\pi} (-\frac{1}{{\bar \tau}}
{\tilde F}^{(a)+}_{mn}{\tilde F}^{(a)+mn}+\frac{1}{\tau}
{\tilde F}^{(a)-}_{mn}{\tilde F}^{(a)-mn}+
2 R^{ab}_{mn}(\,^*{\tilde F})\partial_q\,^*{\tilde F}^{(a)qm}
\partial_p\,^*{\tilde F}^{(b)np}),
\ee
where ${\tilde F}$ are arbitrary two forms in the manifold
and $R$ is the inverse of ${\rm ad}\,^*{\tilde F}$ and it
is a well defined matrix for arbitrary ${\tilde F}$ in
four dimensions.

It can be seen that (\ref{cor9}) is generated from the original
theory by the following
canonical transformation: 
\bea
\label{nonab1}
&&\Pi^{a\alpha}=-4\,^*{\tilde F}^{(a)}_{0\alpha}\nonumber\\
&&A^a_\alpha=\frac14 {\tilde \Pi}^a_{0\alpha}
\eea
where:
\be
\label{nonab2}
{\tilde \Pi}^a_{0\alpha}\equiv\frac{\delta{\tilde L}}
{\delta \,^*\dot{\tilde F}^{(a)}_{0\alpha}}=
4R^{ab}_{\alpha n}\partial_m\,^*{\tilde F}^{(b)}_{nm}.
\ee
The generating functional is given by:
\be
\label{nonab21}
{\cal F}=-4\int_{M_3}{\rm Tr}
(\,^*{\tilde F}_{0\alpha}A_\alpha),
\ee
which has the same form than (\ref{bru5}) but now 
${\tilde F}$ is not derived from a vector potential.
Written in configuration space (\ref{nonab1}) amounts to:
\bea
\label{nonab3}
&&({\bar \tau}-\tau)F^{(a)}_{0\alpha}+
({\bar \tau}+\tau)\,^*F^{(a)}_{0\alpha}=
-2\,^*{\tilde F}^{(a)}_{0\alpha}\nonumber\\
&&A^a_\alpha=R^{ab}_{\alpha n}
\partial_m\,^*{\tilde F}^{(b)}_{nm}.
\eea
Compatibility of these equations is guaranteed on-shell.
The equations of motion in the initial theory
are mapped to the identities:
\be
\label{nonab4}
\partial_n\,^*{\tilde F}^{(a)}_{mn}-
f_{abc}\,^*{\tilde F}^{(c)}_{mn}R^{bd}_{\alpha q}
\partial_p\,^*{\tilde F}^{(d)}_{qp}=0
\ee
in the dual, which are verified straightforwardly
from the second equation in (\ref{nonab3}).
The complete correspondence between equations of
motion and Bianchi identities requires also 
$A^a_0=R^{ab}_{0n}\partial_m\,^*{\tilde F}^{(b)}_{nm}$.
The Bianchi identities in the dual theory are then
satisfied with a ``vector potential'' 
$V^a_n\equiv R^{ab}_{nm}\partial_p\,^*{\tilde F}^{(b)}_{mp}$
defined from the fundamental 2-forms of the dual theory,
and which transforms as a conexion under the local gauge symmetry
of the dual Lagrangian\footnote{${\tilde L}$ has this symmetry
up to a total derivative.}:
${\tilde F}\rightarrow h{\tilde F}h^{-1}$. 
Inversely, the dual equations of motion are mapped to
identities in the original theory.

The initial and dual theories are proved to be equivalent
under (\ref{nonab1}) on-shell.
It would be very interesting to extend the previous formulation
to four dimensional supersymmetric non-abelian gauge theories, 
in particular to
$N=4$ and those $N=2$ for which S-duality is known to be an exact
symmetry \cite{VW} \cite{SW2}. For these theories 
a path integral or a canonical transformation description 
is not known and it could be interesting to
see if such a description exists already at the classical level.
For $N=2$ Yang-Mills theories without matter the low energy
effective action is invariant under S-duality \cite{SW}. In the 
low energy limit the whole non-abelian symmetry group is broken
to its maximal abelian subgroup and then the remaining gauge 
symmetry is abelian. 
The canonical transformation description can then also
be applied straightforwardly to this case.

\subsection*{Acknowledgements}

I would like to thank J. Borlaf for collaboration on some
of the results presented. Also
I am grateful to the organizers of the Argonne Duality
Institute for their warm hospitality and for financial
support.
Work supported by the European Commission TMR programme
ERBFMRX-CT96-0045.

\end{document}